\documentclass[aps,prb,twocolumn,showpacs,groupedaddress]{revtex4}
\usepackage{bm,amsmath,amssymb,latexsym}
\usepackage{graphicx}

\renewcommand{\vec}[1]{\bm{#1}}

\begin{document}

\title{Vector chiral order in frustrated spin chains}

\author{I. P. McCulloch}
\thanks{Current address: Department of Physics, University of Queensland,
Brisbane QLD 4072, Australia}

\author{R. Kube}
\author{M. Kurz}
\author{A. Kleine}
\author{U. Schollw\"ock}

\author{A. K. Kolezhuk}
\thanks{On leave from: Institute of Magnetism, National Academy of Sciences and Ministry
  of Education,  03142 Kiev, Ukraine}

\affiliation{Institut f\"ur Theoretische Physik C, RWTH Aachen, 
D-52056 Aachen, Germany}

\date\today

\begin{abstract}
By means of a numerical analysis using a non-Abelian symmetry realization of the
density matrix renormalization group, we study the behavior of vector chirality
correlations in isotropic frustrated chains of spin $S=1$ and $S=1/2$, subject
to a strong external magnetic field.  It is shown that the field induces a phase
with spontaneously broken chiral symmetry, in line with earlier
theoretical predictions. We present results on the field dependence of the
order parameter and the critical exponents.
\end{abstract}


\maketitle

%%%%%%%%%%%%%%%%%%%%%%%%%%%%%%%%%%%%%%%%%%%%%%%%%%%%%%%%%%%%%%%%%%%%%%

\section{Introduction}

%%%%%%%%%%%%%%%%%%%%%%%%%%%%%%%%%%%%%%%%%%%%%%%%%%%%%%%%%%%%%%%%%%%%%%

The so-called vector chirality in quantum spin chains is defined as the vector
product of two adjacent spins along the chain:
\[
\vec{\kappa}_{n} = \langle \vec{S}_{n}\times\vec{S}_{n+1}
\rangle.
\] 
In a chirally ordered state spins have a tendency to ``rotate'' in some
preferred plane in a certain preferred direction (clockwise or
counterclockwise). Chiral phases in quantum spin chains were predicted long ago
\cite{Villain78,Chubukov91} and have attracted  considerable interest recently
\cite{Nersesyan+98,Kaburagi+99,K00,Lecheminant+01,Hikihara+01,KV05} after they
have been found numerically in frustrated chains with easy-plane anisotropy.
\cite{Kaburagi+99,Hikihara+01} As noted by Villain, \cite{Villain78} the chiral
order should survive at finite temperature in the presence of three-dimensional
interactions without tranforming into a usual helical long-range order: at
finite temperatures the chirality correlation length is much larger than the
spin correlation length, so with decreasing temperature chiral order should
set in before spin order does; there are experimental indications that the
chiral order may exist in the quasi-one-dimensional anisotropic organic magnet
$\rm Gd(hfac)_{3}NITiPr$.\cite{Affronte+99} 
The projection of the vector chirality
$\vec{\kappa}$ on the direction of the applied field could be experimentally detected
with the help of polarized neutrons. \cite{Maleyev+98}

In all known cases of numerically confirmed existence of chiral states, the
preferred plane for spin rotation is chosen by some anisotropy of the easy-plane
type, and the chiral phase disappears in the isotropic limit.\cite{Hikihara+01}
Recently, it has been predicted \cite{KV05} that in isotropic frustrated chains
the chiral phase may appear in presence of an external magnetic field, strong
enough to close the spectral gap. In such a
state, the system approximately decouples into a gapped antisymmetric sector and
a gapless symmetric sector, the latter being described by the Tomonaga-Luttinger liquid (TLL). 
An alternative two-component
TLL scenario \cite{FathLittlewood98,Okunishi+99,OkunishiTonegawa03}
assumes the existence of the Tomonaga-Luttinger liquid in both sectors and implies absence
of the chiral order. 

The phase diagrams of the antiferromagnetic zigzag spin chains in applied field
 have been studied numerically in
Ref.\ \onlinecite{OkunishiTonegawa03} for $S=1/2$, and in Ref.\
 \onlinecite{Heidrich-Meisner+07} for $S=1$. However, both works focused only on the
magnetization process and did not check the presence of the chiral order.
The theoretical analysis of Ref.\ \onlinecite{KV05}
involves some uncontrolled approximations (mean-field decoupling of the
``twist'' term) and, to our knowledge, the chiral order in isotropic spin
chains has never been directly probed numerically (the only exception being the calculation of
 short-range correlations in a $S=1/2$ chain \cite{Yoshikawa+04}), so the question of the correct
scenario remains unsettled.

Several materials are
known which realize isotropic zigzag spin chains; \cite{Hase+04} among them,
$\rm (N_{2}H_{5})CuCl_{3}$ can be viewed as a promising candidate for
experimental studies, since its small exchange constants make it feasible to
reach magnetic fields comparable to the gap.

In this paper we present a study of  vector chirality correlations in the isotropic
$S=1$ and $S=1/2$ zigzag chains in the presence of applied magnetic field, using a powerful
non-Abelian symmetry realization \cite{McCullochGulacsi02} of the density matrix renormalization group
technique \cite{White92,Schollwock-RMP-05} in its matrix product state formulation.\cite{McCulloch07} 
It is demonstrated that the chiral order does exist in the
high-field phase, both for 
$S=1$ and $S=1/2$,
and the behavior of chiral correlations is in a qualitative
agreement with the expectations following from the theoretical analysis of Ref.\
\onlinecite{KV05}. This implies that a
chiral one-component Tomonaga-Luttinger liquid scenario is realized.

%%%%%%%%%%%%%%%%%%%%%%%%%%%%%%%%%%%%%%%%%%%%%%%%%%%%%%%%%%%%%%%%%%%%%%

\section{Theoretical estimates}  

%%%%%%%%%%%%%%%%%%%%%%%%%%%%%%%%%%%%%%%%%%%%%%%%%%%%%%%%%%%%%%%%%%%%%%

We consider  the model of a frustrated antiferromagnetic spin chain, defined by the Hamiltonian:
\begin{equation}
\label{S-ham}
{\mathcal H}=J_{1}\sum_{n} \vec{S}_{n}\cdot \vec{S}_{n+1} 
+J_{2}\sum_{n}\vec{S}_{n}\cdot \vec{S}_{n+2} -H \sum_{n}S_{n}^{z}
\end{equation}
where $\vec{S}_n$ are spin-$S$ operators at the $n$-th site,  
$J_{1}>0$ and $J_{2}>0$ are the nearest and next-nearest neighbor exchange
constants, respectively, and $H$ is the external magnetic field, assumed to
be applied along the $z$ axis.

In case of $S=1$, at $H=0$ the ground state is always
gapped: for small frustration parameter $\alpha\equiv J_{2}/J_{1}$ one remains in the Haldane phase
characterized by the long-range string order, while for
$\alpha>\alpha_{c}\approx 0.75$ there is a first-order transition into another
gapped state, the so-called ``double-Haldane'' (DH) phase where the string order disappears with a
finite jump, giving way to a more complicated hidden order.\cite{KRS96}
When the applied field exceeds the critical value $H=H_{c}$
($H_{c}$ is obviously equal to the gap at $H=0$),
the system acquires finite magnetization. There is another special field value,
the saturation field
$H_{s}$ above which the spins are fully polarized. 
In the $S=1/2$ case, the ground state at $H=0$ is gapless for $\alpha<
\alpha_{c}\simeq 0.24$, and  $\alpha>\alpha_{c}$ corresponds to a gapped
dimerized phase.

We are interested in the
properties of the partially magnetized state in the field range $H_{c}< H<
H_{s}$, and we assume that $\alpha$ is large enough to make the system gapped at
$H=0$.
In the limit $\alpha\gg 1$ the system may
be viewed as two weakly coupled chains. 
A single $S=1/2$ chain in
external field has been extensively
studied,\cite{Bogoliubov+86,AffleckOshikawa99,Furusaki} 
as well as its $S=1$ counterpart. \cite{KonikFendley02,CamposVenuti+02,Sato06,Friedrich+07} 
Above the first critical field $H_{c1}$
the low-energy physics of a single chain is well described
in terms of the effective Tomonaga-Luttinger liquid (TLL) theory,  described by the
Hamiltonian 
\begin{equation}
\label{SpinChainBosHam}
{\cal H}_{TL}[\theta,\varphi] =  \frac{v}{2}\int dx \, \Big\{\frac{1}{K}(\partial_x \phi)^{2} 
+ K (\partial_x \theta)^{2}\Big\}.
\end{equation}
Here  $K$ is the so-called TLL parameter,  $v\propto J_{2}$ is the Fermi velocity, $\phi$ is the bosonic field
(compactified by the condition $\varphi \equiv \varphi+\sqrt{\pi}$),
and  $\theta$ is its dual satisfying the commutation
relations $[\phi(x),\theta(y)] = i\Theta (y-x)$, where $\Theta(x)$ is the
Heaviside function.

\begin{figure}[tb]
\includegraphics[width=0.25\textwidth]{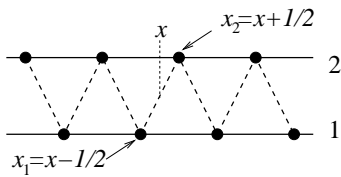}
\caption{\label{fig:zigspin}
A zigzag spin chain and the notation adopted in (\ref{spinops}).}
\end{figure}

In the continuum limit, the lattice spin operators both for $S=1/2$ and $S=1$
can be represented \cite{KV05,Sato06} through the bosonic field $\varphi$ and
its dual $\theta$:
\begin{eqnarray}
\label{spinops}
S^{z}_{a}(x_{a})&=&M+\frac{2}{\sqrt{\pi}} \partial_x \phi_{a}(x_{a})\nonumber\\
&+& A_{3} \sin \big\{\pi M x_{a}+ \sqrt{4\pi }
\phi_{a}(x_{a})\big\} + (\cdots)  \\
S^{+}_{a}(x_{a})& =& e^{i \pi x/2} e^{i\sqrt{\pi}\theta_{a}(x_{a})}
\big\{ A_{1}\nonumber\\
&+&A_{2}\sin{\big(\pi M x_{a}+ \sqrt{4\pi }\phi_{a}(x_{a})\big)}
\big\} + (\cdots),\nonumber
\end{eqnarray}
Here $a=1,2$ labels the two chains, the space coordinate $x$ is defined along
the zigzag path as shown in Fig.\ \ref{fig:zigspin}, the lattice sites
correspond to $x_{1,2}=x\mp \frac{1}{2}$, $M$ is the ground state magnetization
per spin in units of saturation (which corresponds to the filling factor in the
TLL model), and $A_{i}$ are nonuniversal amplitudes.  In the case of $S=1$ the dots
denote additional operators which correspond to massive fields connected to the
high-energy $S^{z}=0,-1$ magnon branches of the Haldane chain.\cite{Sato06}.

The theory parameters $M$, $K$, and $v$ should be
understood as functions of the field $H$; for $S=1$ their behavior 
can be extracted from the comparison of the TLL theory
predictions with the numerical results,\cite{CamposVenuti+02,Fath03}
while for $S=1/2$ it is available from the exact Bethe ansatz slution.\cite{Bogoliubov+86}

For $S=1$ the most important feature \cite{KonikFendley02,CamposVenuti+02,Fath03} is that
generally the TLL parameter $K>1$ at $H>H_{c1}$, non-monotonically depends on $H$  and  tends to the free
fermion value $K=1$ both at the first critical field
$H=H_{c1}$ and at the saturation field $H=H_{s1}$. In contrast to that, for
$S=1/2$ the TLL parameter $K<1$ and is a monotonically increasing function of the
applied field.\cite{Bogoliubov+86,AffleckOshikawa99}

For the description of weakly coupled chains it is convenient to introduce the
symmetric and antisymmetric combinations of the bosonic fields
\[
\phi_{\pm}=(\phi_1\pm \phi_2)/\sqrt{2},\quad   
\theta_{\pm}= (\theta_1 \pm \theta_2)/\sqrt{2}.
\]
The longitudinal ($S^{z}S^{z}$) part of the zigzag exchange, apart from producing
terms of the type
$(\partial_{x}\varphi_{1})(\partial_{x}\varphi_{2})$ which lead to a splitting of the
TLL parameter values for the symmetric and antisymmetric sectors, 
\begin{eqnarray} 
\label{K-pm} 
&& K_{\pm}\approx K\big[1\pm 2K/(\pi v \alpha))\big]^{-1/2}, \nonumber\\
&& v_{\pm}\approx v\big[1\pm 2K/(\pi v \alpha))\big]^{1/2},
\end{eqnarray}
yields another contribution proportional to 
$\cos\big[ \sqrt{8\pi }\varphi_{-}
-\pi M\big]$.  The scaling dimension of this latter perturbation is $2K_{-}$. In
the case of $S=1$ it
is irrelevant since $K>1$ and so can be neglected; in contrast to that, for
$S=\frac{1}{2}$ chain  $K<1$, and this operator is a relevant perturbation. 
Thus, as pointed out in Ref.\
\onlinecite{KV05}, for $S=1$ the leading contribution to the interaction is given by the
``twist term'' produced by the transversal part of the zigzag exchange.
For $S=1/2$ the twist term
competes with a relevant operator and can only win if $K_{-}(H)$ is above a certain
critical value $K_{c}$; for $K<K_{c}$ the so-called ``even-odd'' phase is
realized, whose dominant correlations are of the spin-nematic (or XY2 in the
nomenclature of Ref.\ \onlinecite{Schulz86}) type.
A characteristic feature of the ``even-odd'' phase\cite{OkunishiTonegawa03} are the $\Delta S^{z}=2$
steps in the magnetization curve $S^{z}_{\rm tot}(H)$. The ``even-odd'' phase
has been
also observed\cite{Vekua+07} in zigzag chains with ferromagnetic nearest-neighbor exchange.

In the chirally ordered phase the twist term is the most
relevant perturbation, so one obtains the same effective Hamiltonian  for $S=1$  as well as for  $S=1/2$:
\begin{eqnarray}
{\mathcal H}_{\rm eff}  &=&
\sum_{\sigma=\pm}\mathcal{H}_{TL}[\varphi_{\sigma},\theta_{\sigma}] +
\mathcal{H}_{\rm int}\nonumber\\
\mathcal{H}_{\rm int} &=& g \int dx \sin\big(\sqrt{2\pi} \theta_{-}\big)(\partial_{x}\theta_{+}).
\end{eqnarray}
Mean-field decoupling of the twist term in the spirit of 
Ref.\ \onlinecite{Nersesyan+98} then leads to the conclusion\cite{KV05}
that both $\langle \partial_{x}\theta_{+} \rangle$ and $\langle
\sin\big(\sqrt{2\pi} \theta_{-}\big) \rangle$ become nonzero, and the antisymmetric
sector becomes gapped. 

One should mention that the above description makes sense only when we are far
enough from the critical fields $H_{c}$ or $H_{s}$: the theory is applicable only up
to the energies of the order of the bandwidth $v$, and $v\to 0$ if $H\to
H_{c},H_{s}$. The formulae (\ref{K-pm}) indicate that the system becomes
unstable against phase separation\cite{CazalillaHo03}
as soon as $v<2K/(\pi\alpha)$.

\begin{figure}[tb]
\includegraphics[width=0.42\textwidth]{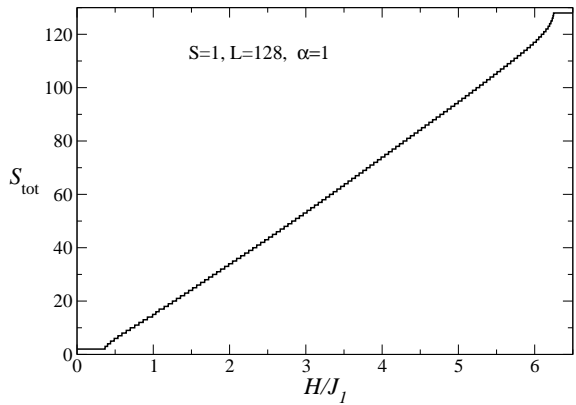}
\caption{\label{fig:magS1} Numerically calculated magnetization curve
  $S^{z}_{\rm tot}(H)$ for a $S=1$  zigzag chain of  length  $L=128$ with 
  frustration parameter $\alpha=J_{2}/J_{1}=1$. 
}
\end{figure}

The components of the chirality operator $\vec{\kappa}$ can be expressed through
bosonic fields. The longitudinal part of the chirality
can be obtained in the following form:
\begin{eqnarray} 
\label{kappa-z} 
\kappa^{z}(x)&=& \sin\big(\sqrt{2\pi} \theta_{-}\big) \Big\{ A_{1}^{2}-
\frac{(\pi A_{1})^{2}}{4}(\partial_{x}\theta_{+})^{2} \nonumber\\
&+&\frac{A_{2}^{2}}{2}\cos\big(\sqrt{8\pi}\varphi_{+} +2\pi M x\big) \Big\} + (\cdots),
\end{eqnarray}
where dots denote massive fields (the most important contribution of that sort
is proportional to $(-1)^{x} \cos\big(\sqrt{2\pi}
\theta_{-}\big)(\partial_{x}\theta_{+})$ ) and operators with higher scaling
dimensions. The leading contribution to the long-distance correlator is thus
given by
\begin{equation} 
\label{kappa-z-corr} 
\langle \kappa^{z}(x)\kappa^{z}(0)\rangle \to \kappa_{0}^{2}\Big(1+ \frac{C_{1}}{x^{4}}
+\frac{C_{2}\cos(2\pi Mx)}{x^{4K+}}\Big) ,\quad x\gg \xi,
\end{equation}
where $\xi$ is the largest correlation length determined by the gap in the
antisymmetric sector. (For $S=1$ there is also another, much smaller, characteristic
correlation length  $\widetilde{\xi}$ which is determined by the high-lying excitation branches that are
neglected in the bosonization formulae (\ref{spinops}); it roughly corresponds
to the correlation length of the Haldane chain at zero field, typically a few
lattice constants.)
For $S=1$, although $K>1$ for $H_{c}< H < H_{s}$, the parameter $K_{+}$, according to (\ref{K-pm}), is renormalized to
smaller values when the zigzag coupling is switched on, so the two decaying
contributions in (\ref{kappa-z-corr}) may be competing with each other. In
$S=1/2$ chain the oscillating contribution always has the slowest decay since $K_{+}<K<1$.

\begin{figure}[tb]
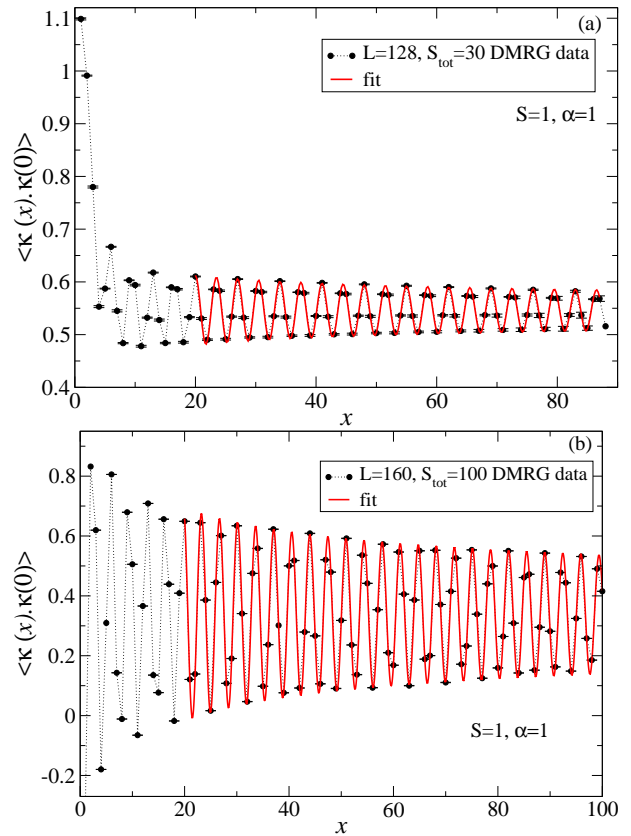

\includegraphics[width=0.45\textwidth]{F128-30.eps}
\includegraphics[width=0.45\textwidth]{F160-100.eps}
\caption{\label{fig:fits}
(Color online) 
Typical DMRG results for the chirality correlator (\ref{mixed-corr}) and results of fitting it to the  form (\ref{fit-law}).
The error bars\cite{note1} (here smaller than the symbol size) indicate the variance of $F(x=n-n')$ calculated from averaging over 
the initial and final points $n$,
$n'$.
}
\end{figure}

In a similar way one can obtain the transversal chirality component
\begin{eqnarray} 
\label{kappa-plus} 
\kappa^{+}(x) &=& 2A_{1}M \sin\big(\sqrt{\pi/2} \theta_{-}\big) \nonumber\\
&\times& \exp\Big\{ \frac{i\pi x}{2} +i\sqrt{\frac{\pi}{2}}\theta_{+} \Big\} + (\cdots).
\end{eqnarray}
It is easy to see that the leading term in $\kappa^{+}(x)$ is simply
proportional to $S_{1}^{+}-S_{2}^{+}$.
The leading contribution to the corresponding asymptotic correlator is
slowly decaying
\begin{equation} 
\label{kappa-plus-corr} 
\langle \kappa^{+}(x)\kappa^{-}(0)\rangle \propto A_{1}^{2} 
\frac{ M^{2}}{x^{1/(4K_{+})}} \exp\{i Q x\}, \quad x\gg \xi,
\end{equation}
and incommensurate, with the wave vector given by 
\[
Q=\frac{\pi}{2} +\sqrt{\frac{\pi}{2}}\langle \partial_{x}\theta_{+}\rangle
\]

The above expressions (\ref{kappa-z-corr}) and (\ref{kappa-plus-corr}) are
expected to be valid in the limit $\alpha\gg1$, and for $H$ not very close to
the critical fields $H_{c}$, $H_{s}$ (in the vicinity of the critical field the
bosonization approach becomes hardly applicable since the effective bandwidth
goes to zero). Close to the saturation field $H_{s}$, a large-$S$ analysis 
\cite{KV05} allows mapping the system to an effective model of two bosonic
species with repulsive interaction, which condense driven by the magnetic field
playing the role of the chemical potential. The repulsion turns out to be strong
enough to satisfy the phase separation condition, so only one of the species
condenses,  the other condensate is depleted, so one deals in fact with the
one-component pseudo-condensate whose physics is again described by a
(one-component) TLL. The asymptotic form of
the longitudinal chirality correlator for $H$ close to $H_{s}$ has been
presented in Ref.\ \onlinecite{KV05}:
\begin{equation} 
\label{kappa-z-corr-Hs} 
\langle \kappa^{z}(x)\kappa^{z}(0)\rangle \to  \kappa_{0}^{2} - \frac{C}{x^{2}},
\end{equation}
with $\kappa_{0}^{2}\propto (H_{s}-H)$.  The leading contribution to the
transversal chirality is proportional to the bosonic field itself, so its
correlator takes the following asymptotic form:
\begin{equation} 
\label{kappa-plus-corr-Hs}
\langle \kappa^{+}(x)\kappa^{-}(0)\rangle \to \frac{C'}{x^{1/(2K')}}  e^{i Q' x}.
\end{equation}
Here $K'$ is another TLL parameter, the 
characteristic wave vector $Q'$ is given by the expression for the
pitch of the classical helical state:
\begin{equation} 
\label{pitch} 
Q'=\pm(\pi -\arccos(1/4\alpha)),
\end{equation}
and the amplitude
$C'\propto (H_{s}-H)^{1/2-1/(4K')}$. As the field approaches the saturation
point, $H\to H_{s}$, the value of $K'$ tends to
 $1$, so the amplitude
$C'$ vanishes.

%%%%%%%%%%%%%%%%%%%%%%%%%%%%%%%%%%%%%%%%%%%%%%%%%%%%%%%%%%%%%%%%%%%%%%
\section{Results of numerical analysis}

%%%%%%%%%%%%%%%%%%%%%%%%%%%%%%%%%%%%%%%%%%%%%%%%%%%%%%%%%%%%%%%%%%%%%%

We have studied the $S=1$ and $S=1/2$ zigzag chain model given by (\ref{S-ham}) using the
DMRG method  in its matrix product state formulation, making full use  of the
non-Abelian $SU(2)$ symmetry.

For
a full description of the DMRG technique,\cite{White92}, we refer the reader to
the review Ref.\ \onlinecite{Schollwock-RMP-05}. As discussed in Ref.\
\onlinecite{McCulloch07}, the formulation in terms of matrix product states is
very convenient but for the calculation of ground states does not lead to
substantially better results.  The decisive point
\cite{McCullochGulacsi02,McCulloch07} is the use of the non-Abelian symmetry
$SU(2)$ instead of the Abelian $U(1)$. While the magnetic field $H$ breaks
$SU(2)$ symmetry, the fact that the Zeeman energy term commutes with the rest of
the Hamiltonian makes it possible to take take the influence of the magnetic
field into account by calculating the ground state of the model in a sector with
the given total spin $S_{\rm tot}$. The advantage of the method lies in a
drastic reduction of the number of states $m$ which is necessary to describe the
system, because non-Abelian symmetry allows one to calculate using
representatives of groups of states of the same total spin: essentially, one
treats the multiplet of states of the same total spin as a single representative
state.
Comparing to the Abelian version of the method which only uses the $U(1)$
symmetry, the improvement in efficiency can be several orders of magnitude,
depending on the problem. For the zero-field groundstate of the 
zigzag chain, the effective improvement 
in the number of states is a factor $\sim 3$ (for $S=1/2$) or $\sim 4$ (for $S=1$),
leading to a reduction in the computational effort by a factor $\sim 27$ and $\sim 64$ 
respectively. The relative efficiency decreases as the magnetic field is increased,
but even for rather high fields the improvement is appreciable.

\begin{figure}[tb]
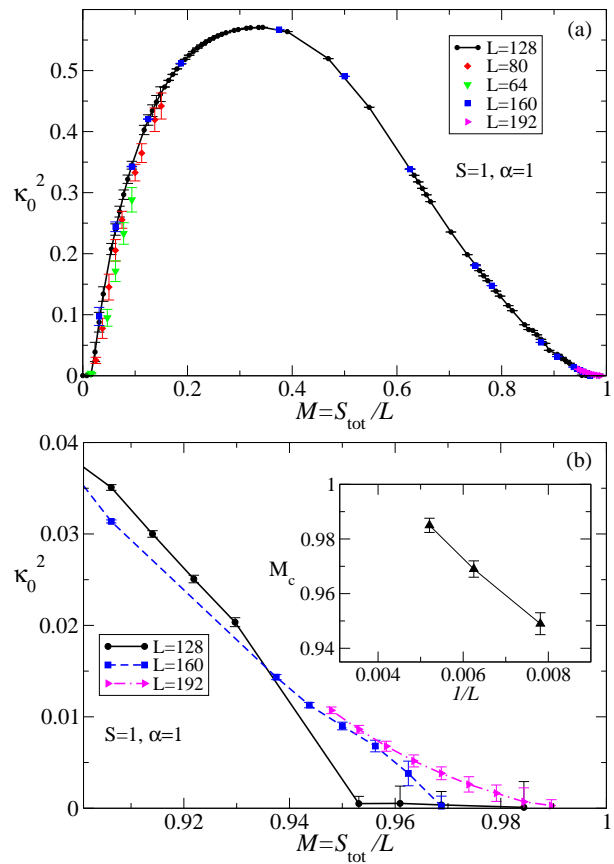

\includegraphics[width=0.45\textwidth]{kappaM-a.eps}
\includegraphics[width=0.45\textwidth]{kappaM-b.eps}
\caption{\label{fig:kappaM}
(Color online)
(a) the square chirality order parameter $\kappa_{0}^{2}$ as a function of
magnetization $M=S_{\rm tot}/L$ for a
$S=1$ zigzag chain with $\alpha=1$, extracted from fits of chirality correlation
functions; 
(b) zoom-in of the same picture in the vicinity of $M=1$, where finite size effects become
important. The inset shows the $L$ scaling of the point $M_{c}$ where the long-range
chirality order disappears in a finite system of size $L$.
}
\end{figure}

A slight disadvantage is that the non-Abelian method allows to compute only reduced
matrix elements (in the sense of the Wigner-Eckart theorem). 
In our case, since the chirality is a vector, the correlator that is by far the easiest
to calculate is the rotationally invariant scalar product,
\begin{equation} 
\label{mixed-corr} 
F(n-n')=\langle \vec{\kappa}(n)\cdot \vec{\kappa}(n')\rangle,
\end{equation}
which is a mixture of the longitudinal and transversal contributions.\cite{note2}
This obviously makes the
analysis of the numerical data more
difficult: in our case, from the theoretical analysis it follows that the
longitudinal chirality correlations
$\langle \kappa^{z}(x)\kappa^{z}(0)\rangle$  decay to their asymptotic value
much faster than the transversal ones. Thus it turns out to be practically impossible to extract the
characteristic decay exponent $\eta_{z}$ for the longitudinal chirality correlations from
the $F(x)$ data, and one can only try to estimate the exponent $\eta=\eta_{xy}$ of
the transversal chirality correlations.

\begin{figure}[tb]
\includegraphics[width=0.45\textwidth]{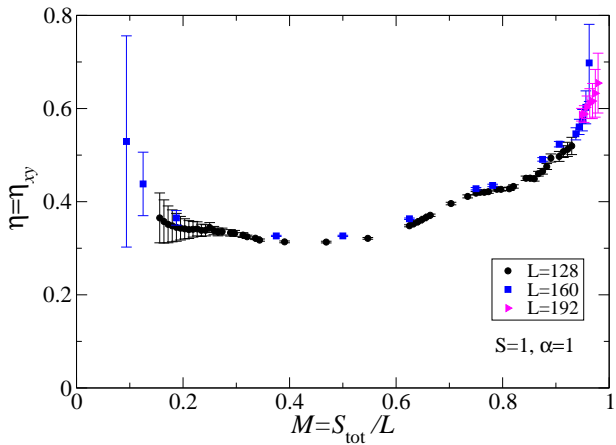}
\caption{\label{fig:etaM} (Color online) Behavior of the transversal chirality correlations
exponent $\eta$ for a $S=1$ chain with $\alpha=1$ as a function of magnetization $M=S_{\rm tot}/L$,
extracted from fits of the correlation function (\ref{mixed-corr}) to the
functional form (\ref{fit-law}). The error bars shown correspond to the
uncertainties of the fit. 
}
\end{figure}

%%%%%%%%%%%%%%%%%%%%%%%%%%%%%%%%%%%%%%%%%%%%%%%%%%%%%%%%%%%%%%%%%%%%%%%%%%%%%%%%

\subsection{$S=1$ zigzag chain}

We have 
studied
spin-$1$ zigzag chains with the
frustration parameter $\alpha=1$, for several chain lengths $L$ ranging from
$64$ to $192$. 
For our calculation, even within the $SU(2)$ method we needed a
relatively large number of representative states(from $m\simeq 400$ to $m\simeq 2000$,
depending on $L$ and $S_{\rm tot}$) to reach 
good convergence.  
In the $U(1)$ formulation, this corresponds to an $m$ of up to $8000$,
making an Abelian calculation much more difficult.

As one can see from the numerically calculated magnetization  curve shown in Fig.\ \ref{fig:magS1},
at $\alpha=1$ the $S^{z}(H)$ dependence is featureless and shows neither plateaux,
nor cusps, nor $\Delta S^{z}=2$ steps characteristic for even-odd phase,
in accordance with the results of Ref.\ \onlinecite{Heidrich-Meisner+07}.

\begin{figure}[tb]
\includegraphics[width=0.45\textwidth]{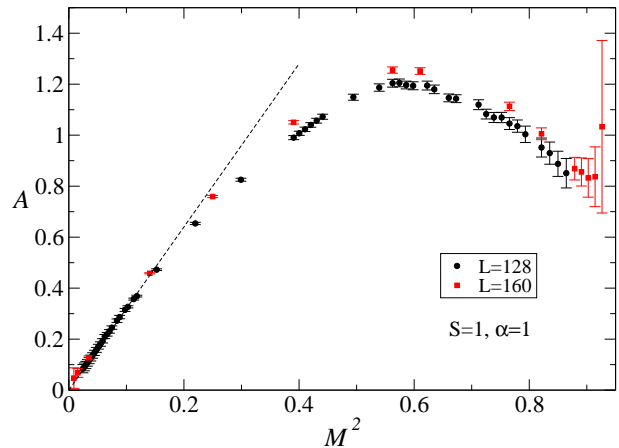}
\caption{\label{fig:A-M} (Color online) Behavior of the oscillations amplitude $A$  as a function of
the magnetization $M=S_{\rm tot}/L$ for a $S=1$ chain with $\alpha=1$. One can see that the scaling $A\propto
M^{2}$ suggested by (\ref{kappa-plus-corr}) is only applicable for
small $M$.
}
\end{figure}

We have computed the chirality correlator (\ref{mixed-corr}) in the ground
states of a large number of sectors with certain total spin quantum number $S_{\rm tot}$.
When
computing $F(n-n')$, it was averaged over the starting and final positions $n$,
$n'$, and care was taken to stay in the bulk, away from the ends of the chain.
The DMRG data for the correlator has been fitted to the power-law form
\begin{equation} 
\label{fit-law} 
F(x) =\kappa_{0}^{2}+ \frac{A\cos[q(x+\delta)]}{x^{\eta}}
\end{equation}
suggested by (\ref{kappa-z-corr}), (\ref{kappa-plus-corr}).
The introduction of a finite phase shift $\delta$ is necessary due to the open
boundary conditions. Typical fits are presented in Fig.\ \ref{fig:fits}.
From those fits we have extracted the behavior of the
chirality order parameter $\kappa_{0}^{2}$ and the exponent $\eta$ as functions
of the chain magnetization $M=S_{\rm tot}/L$, shown respectively 
in Fig.\ \ref{fig:kappaM} and Fig.\ \ref{fig:etaM}. 
The fitted  wave vector $q$  only weakly depends on the magnetization: as $M$
changes from $0$ to $1$, $q$ varies from $1.79$ to $1.83$, which favorably
compares to the classical value  $Q'\approx 1.82$ obtained from (\ref{pitch}) at
$\alpha=1$. The phase shift $\delta(\alpha=1)\approx 1\pm 0.05$ is also
practically independent of $M$. The behavior of the oscillation amplitude is
shown in Fig.\ \ref{fig:A-M}: the scaling $A\propto M^{2}$ suggested by
(\ref{kappa-plus-corr}) is indeed observed for small $M$, and strong deviations
appear for $M>0.3$.

One can see that for the bulk of $M$ values the order parameter $\kappa_{0}^{2}$
has practically reached convergence already at $L=128$, so there is no need to
perform the finite-size scaling.  The only region where finite-size effects
remain strong is $M\to 1$ (see Fig.\ \ref{fig:kappaM}b): the
$\kappa_{0}^{2}(M)$ dependence at finite $L$ shows $\kappa_{0}$ vanishing at
some $M=M_{c}\not=1$. The finite-size scaling of $M_{c}$ (see the inset of Fig.\
\ref{fig:kappaM}b) shows that there is no trend to convergence even for $L=192$,
although there is a sizeable increase in $M_{c}$ towards $1$ with increasing the
size $L$. From the theoretical analysis one expects $M_{c}\to 1$ for $L\to
\infty$, however, studying this limit numerically can be quite difficult: since
for $M\to 1$ one is very close to the fully polarized state, the actual number
of particles in the problem is the number of magnons $L(1-M)$, which has to be
large enough to observe a phase transition.

The quality of fits deteriorates for very large $S_{\rm tot}$, since
the overall scale of the chirality correlations goes to zero as $M\to 1$.
The fits become less reliable for small $S_{\rm tot}$ as well, for the following reason:
with (\ref{fit-law}) one attempts to fit just the oscillating (transversal) part
of the chirality correlator. For small $S_{\rm tot}$ the amplitude of the
oscillating part is small, while the gap in the antisymmetric sector is small,
so the oscillations appear on top of the exponential decay characterized by two
very different correlation lengths $\widetilde{\xi}$ (for $\alpha=1$ one has
$\widetilde{\xi}\approx 7$) and $\xi\gg \widetilde{\xi}$. It thus becomes a
numerically ill-posed problem to filter out the power-law decaying oscillating
part on top of such a background.

\begin{figure}[tb]
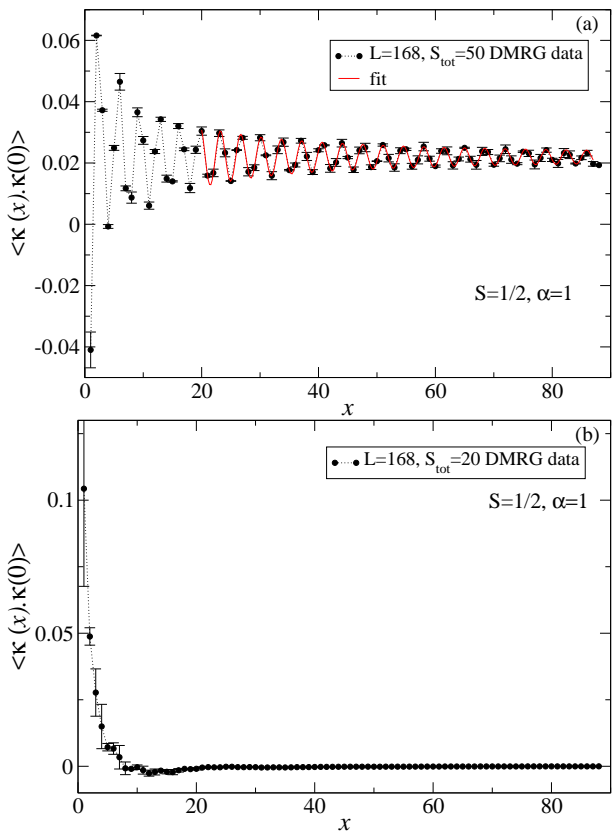

\includegraphics[width=0.45\textwidth]{F-oh-L168-S50.eps}
\includegraphics[width=0.45\textwidth]{F-oh-L168-S20.eps}
\caption{\label{fig:fits-oh}
Typical DMRG results for the chirality correlator (\ref{mixed-corr}) of a
$S=1/2$ zigzag chain with $\alpha=1$. 
The error bars have the same meaning as in Fig.\ \ref{fig:fits}.
(a) (Color online) A point in the chiral phase. The solid line shows a fit  to (\ref{fit-law});
(b) a correlator in the nonchiral ``even-odd'' phase.
}
\end{figure}

The extracted critical exponent slowly changes with $M$ and lies in the range
$\eta\sim 0.3\div 0.6$, which qualitatively agrees with the theoretical
estimates predicting that for $\alpha\gg 1$ it should vary from approximately
$0.25$ to $0.5$; the error bars shown in Fig.\ \ref{fig:etaM}) are in fact
only of indicative nature since they only show uncertainties of the fit to the fixed fit
function and do not take into account the variations of the fit parameters which
would result from adding subleading (e.g., exponentially decaying) contributions
to Eq.\ (\ref{fit-law}).

The chosen value of $\alpha=1$ is rather small and does not allow a direct
comparison of $\eta$ with the theoretical value $(4K_{+})^{-1}$:
if one naively tries to extract the velocity parameter $v$ assuming that
 $K_{+}$ is given by
(\ref{K-pm}) and using the data of Ref.\ \onlinecite{Fath03} for the $K(M)$
dependence, the obtained values of $v$ fall below the phase separation threshold
$v_{c}=2K/(\pi\alpha)$. We have refrained from studying chains with $\alpha\gg
1$ since, on the one hand, increasing $\alpha$ causes a dramatic increase in
numerical effort, and on the other hand, the chirality correlators become more
and more ``polluted'' by the exponentially decaying contributions with ever larger
correlation length $\xi$.

\begin{figure}[tb]
\includegraphics[width=0.48\textwidth]{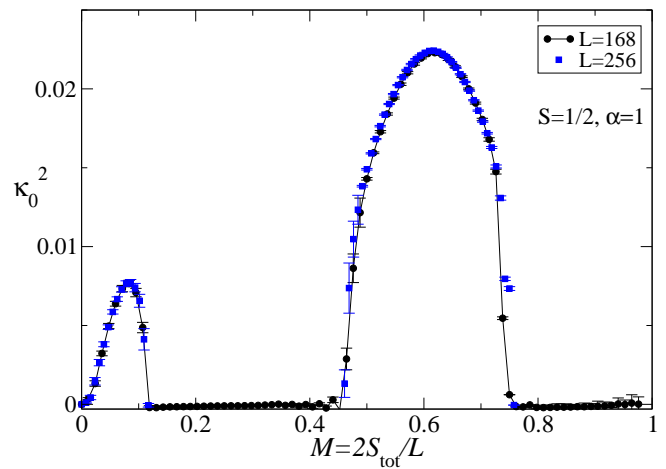}
\caption{\label{fig:kappa-oh-M}
(Color online) Behavior of the square chirality $\kappa_{0}^{2}$ as the function of
magnetization $M=2S_{\rm tot}/L$ for a
$S=1/2$ zigzag chain with $\alpha=1$, extracted from fits of chiral correlation
functions. 
}
\end{figure}

\begin{figure}[tb]
\includegraphics[width=0.48\textwidth]{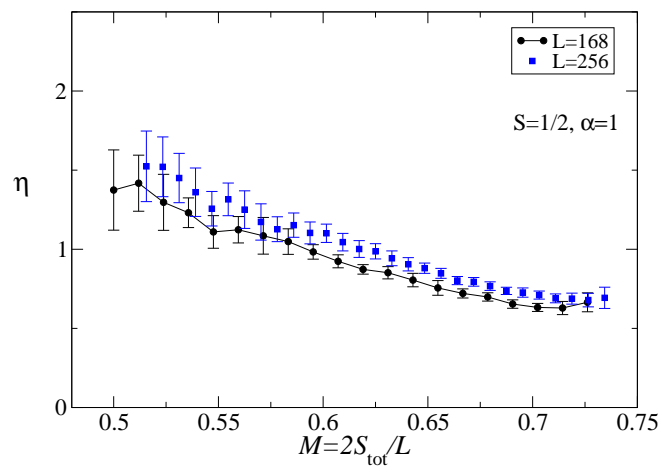}
\caption{\label{fig:eta-oh-M} (Color online) Behavior of the transversal chirality correlations
exponent $\eta$ for a $S=1/2$ chain with $\alpha=1$ as the function of magnetization $M=2S_{\rm tot}/L$,
extracted from fits of the correlation function (\ref{mixed-corr}) to the
functional form (\ref{fit-law}). The error bars shown correspond to the
uncertainties of the fit. 
}
\end{figure}

%%%%%%%%%%%%%%%%%%%%%%%%%%%%%%%%%%%%%%%%%%%%%%%%%%%%%%%%%%%%%%%%%%%%

\subsection{$S=1/2$ zigzag chain}

We have also computed the chirality correlation function (\ref{mixed-corr}) for
$S=1/2$ zigzag chains of length $L=168$, $256$, with  frustration
parameter $\alpha=1$. Typically, $300$ to $400$ representative $SU(2)$ states
were kept in the calculation. The magnetization curve for a $S=1/2$ zigzag chain with
$\alpha=1$ has been presented in Fig.\ 3a of Ref.\ \onlinecite{OkunishiTonegawa03}.
According to
 the phase diagram obtained in Ref.\ \onlinecite{OkunishiTonegawa03},  at $\alpha=1$ the
 $S=1/2$ chain exhibits several phases with varying the applied field $H$. Some
 of those phases, namely, the ``even-odd'' phase and the plateau phase, obviously do not possess
any  chiral order. ``Suspect'' with respect to chirality are only two regions
marked ``TL2''  in Ref.\
 \onlinecite{OkunishiTonegawa03} and identified as a two-component
 Tomonaga-Luttinger liquid phase. Indeed, we observe a finite value of vector
 chiral order in both TL2 regions (although not in the entire high-field TL2
 piece, see below).

A typical example of the correlator in
the chiral phase is shown in Fig.\ \ref{fig:fits-oh}a; for a comparison, we also
show a correlator in the ``even-odd'' phase which is nonchiral.
The maximal magnitude of chirality in $S=1/2$ chain is roughly one order of
magnitude smaller than in the $S=1$ case. To analyse the chirality correlators,
we had to use relatively large chain lengths, because the gaps are smaller than
in the $S=1$ case, and the results for small $L$
are polluted by slow exponentially decaying contributions.

We have employed the same fitting procedure as descibed above for the $S=1$
 chain, and analyzed the behavior of the
chirality order parameter $\kappa_{0}^{2}$ and the critical exponent $\eta$ as functions
of the  magnetization $M=2S_{\rm tot}/L$. The results are shown respectively 
in Fig.\ \ref{fig:kappa-oh-M} and Fig.\ \ref{fig:eta-oh-M}. For the low-$M$
 chiral region, the amplitude of oscillations in the correlation function turns
 out to be too small to extract the exponent $\eta$ with any reasonable
 accuracy, so for that region we were only able to extract the order parameter $\kappa_{0}^{2}$.

The boundaries of the low-field piece of the chiral phase coincide with the
 low-field TL2 region of Ref.\ \onlinecite{OkunishiTonegawa03}. Surprisingly,
 this is not the case for the high-field piece: while its lower boundary
 reasonably agrees with the transition from ``even-odd'' phase to TL2, its upper
 boundary lies at a finite $M=M_{c}\simeq 0.75$ and \emph{not} at $M=1$ as one
 expects from the theoretical analysis. 
It is worth mentioning that
 the magnetization curve of the $S=1/2$ chain at $\alpha=1$ (see Fig.\ 3a of
 Ref.\ \onlinecite{OkunishiTonegawa03}) seems to exhibit a weak feature around
 $M\simeq 0.75$, namely a fast growth of the second derivative $d^{2}M/dH^{2}$.
At $M\to M_{c}$ the critical exponent
 $\eta$ tends to $1/2$, the value which is expected theoretically close to the
 saturation field.

In contrast to $S=1$, where the
 respective boundary exhibited strong finite-size scaling, in the $S=1/2$ case
 we have not observed any significant change of $M_{c}$ with increasing $L$ from
 $168$ to $256$, as seen from Fig.\ \ref{fig:kappa-oh-M}. We have found no
 chiral order for $L=516$ chain with $S_{\rm tot}=205$, which means that even
 for such a long chain $M_{c}(L=516)< 0.787$.
On the basis of
 available data, one can conclude that the $S=1/2$ chain might possess another
 nonchiral phase close to the saturation field. 
The nature of this 
 phase needs further investigation.

\section{Summary} 

We have studied spin-$1$  and spin-$1/2$ isotropic
antiferromagnetic zigzag chains in strong magnetic fields by means  of the matrix
product density matrix renormalization group technique.
Existence of  a phase with field-induced vector chiral order is established for
$S=1$ as well as for $S=1/2$, and the
behavior of the order parameter and its correlations as functions of the
magnetization is analyzed.   The chiral phase is
\emph{gapless} and corresponds to a
one-component Luttinger liquid, thereby confirming the scenario proposed in
Ref.\ \onlinecite{KV05}.

\begin{acknowledgments} 

We would like to thank T. Vekua 
for useful discussions.
This work was partly supported by the
Deutsche Forschungsgemeinschaft (DFG).
A.K.K. was supported by the Heisenberg Program Grant No.\
KO~2335/1-2 from DFG.

\end{acknowledgments}

%%%%%%%%%%%%%%%%%%%%%%%%%%%%%%%%%%%%%%%%%%%%%%%%%%%%%%%%%%%%%%%%%%%%%%%%%%


\begin{thebibliography}{10}
\bibitem{note1} In that way, the error bars shown do not reflect the
  actual error of the DMRG calculation, which is much smaller, but merely
  reflect the strength of finite-size effects.
\end{thebibliography}
\end{document}